# Uncovering Volatility Dynamics in Daily REIT Returns[*]

John Cotter, University College Dublin[†]
&
Simon Stevenson, Cass Business School, City University and
University College Dublin[‡]

[†] Centre for Financial Markets, Department of Banking & Finance, University College Dublin, Blackrock, County Dublin, Ireland, E-Mail: john.cotter@ucd.ie

[‡] Real Estate Finance & Investment Group, Faculty of Finance, Cass Business School, City University, 106 Bunhill Row, London, EC1Y 8TZ, UK.
Centre for Real Estate Research, Smurfit School of Business, University College Dublin, Blackrock, County Dublin, Ireland, Tel: +353-1-7168848, Fax: +353-1-2835482,
E-Mail: simon.stevenson@ucd.ie

[*] The authors would like to thank participants at the 2nd Annual Research Symposium of the Centre for Real Estate Research, University College Dublin and the 2004 American Real Estate Society Annual Meeting for comments on previous drafts of this paper.

# Uncovering Volatility Dynamics in Daily REIT Returns

## Abstract

Using a time-varying approach, this paper examines the dynamics of volatility in the REIT sector. The results highlight the attractiveness and suitability of using GARCH based approaches in the modeling of daily REIT volatility. The paper examines the influencing factors on REIT volatility, documenting the return and volatility linkages between REIT sub-sectors and also examines the influence of other US equity series. The results contrast with previous studies of monthly REIT volatility. Linkages within the REIT sector and with related sectors such as value stocks are diminished, while the general influence of market sentiment, coming through the large cap indices is enhanced. This would indicate that on a daily basis general market sentiment plays a more fundamental role than more intuitive relationships within the capital markets.

# Uncovering Volatility Dynamics in Daily REIT Returns

**Introduction**

Recent years have seen a structural change in the attitude of investors to real estate securities. Much of this shift has accompanied the strong performance in many markets of real estate securities. In addition to the strong recent performance of the sector, there have been a number of structural changes in the sector that have further increased its attractiveness. The growth in the REIT sector in the United States is of particular interest. The REIT structure overcomes many of the limitations in conventional indirect real estate vehicles, in particular the issue of tax transparency. The standard corporate structure used in countries such as the UK provides a disincentive for many institutional investors from holding such securities, and a corresponding advantage in favor of direct investment. The growth in REIT type structures in non-US markets further illustrates the advantage to such structures. In the US, the inclusion of REITs in major indices such as the S&P500 has also increased investor awareness and investment, particularly from index based fund managers. The combination of factors such as this, the limitations on REITs in relation to dividend payments and the strong relative performance of the sector in the aftermath of the collapse of the technology bubble have resulted in increased fund flows into the sector. Ling & Naranjo (2004) illustrate the impact of the flow of funds into REITs and the subsequent impact upon REIT returns.

In addition to returns, investors should also be interested in the volatility of these markets given the risk-return trade-off underpinning the performance of financial markets. The current paper incorporates this motivation by examining the dynamics of daily volatility in the REIT sector. Much of the research concerning REITs has been primarily concerned with either the sectors performance as a portfolio asset or the financial characteristics of REITs. Relatively few studies have examined the relationship between REITs and mainstream capital market assets. However, as the REIT sector has developed, the direct and indirect influence of the broader capital markets is of increasing importance in the context of REITs as broader investor awareness increases. This is particular so in relation to the issue of volatility. This study aims to assess not only the dynamics underlying REIT volatility, but also to examine the influence of other capital market assets on the sector. The development of the sector may result in a changing relationship between REITs and other equity sectors, particularly at higher frequencies of data. Previous studies of REITs have largely used at most monthly data. The use of daily data in this paper is a deliberate choice. While a common criticism of the use of daily data in academic studies is the noise contained within it, increased investment in REITs, particularly from more active investors, is likely to see increased daily trading and therefore daily volatility in the asset. The analysis of volatility also has important implications in terms of issues such as risk management and the implementation of hedging strategies. In relation to risk measures such as value-at-risk, the estimation of volatility is a key element. The impact of increased investor awareness on daily trading will therefore in all likelihood result in the increased need for the accurate assessment of both volatility itself and the broader linkages between REITs and other equity sectors on a daily basis.

While a literature has developed that has examined the linkages with other capital market assets, the majority of papers have examined linkages in the first moment of the return series and examined the issue of integration and segmentation; for example, Liu et al. (1990) cannot reject the hypothesis that REITs are integrated with common equities. Evidence with regard to the integration of REITs and common stocks has also been found in studies such as Mei & Lee (1994) and Li & Wang (1995). Ling & Naranjo (1999) use multi-factor asset pricing techniques to examine whether there is any evidence of integration between direct real estate, REITs and common stocks. As with previous studies, REITs are found to be integrated with non-real estate equities, however, no such evidence is found in relation to the direct market, even when this data is adjusted for smoothing. Wilson & Okunev (1996) examine the Australian, American and British indirect real estate and equity markets, finding in all three markets an absence of any cointegrating relationships. Okunev & Wilson (1996) use a non-linear integration test to examine the relationship between REITs and the S&P 500 Composite. The results show that while the two markets may be related in a non-linear fashion, the level of deviations between the two can be extensive, with the degree of mean reversion quite slow. For example, the authors find that the half-life of deviations is 30 months is some cases.

A number of studies have examined the issue of substitutability between different REIT sectors, with most emphasis placed on the equity and mortgage sectors. Seck (1996) argues that equity and mortgage REITs are not substitutable due to the fact that they respond differently to common factors. Similar evidence is reported by papers such as Peterson & Hsieh (1997) and Glascock et al. (2000). Glascock et al. (2000) report that while the sectors were substitutable prior to 1992, with evidence of cointegration between the two sectors and common driving forces, this affect is not evident in the post-1992 environment. This result is similar to many that have examined REITs, showing that the early nineties saw a turning point in the price behavior of REITs, and in particular Equity REITs. To a large extent this shift was due to the reforms contained in the 1986 Tax Reform Act, which eliminated many of the tax based investment incentives of REITs. Prior to this legislation many REITs had been effectively established as tax shelters. A recent paper by Lee & Chiang (2004) however, finds further evidence of commonalities between the equity and mortgage sectors. The authors use a variance ratio test; with the results supporting the hypothesis that the two sectors are substitutable, even post early nineties. He (1998) finds evidence to support the notion that a causal relationship exists from Equity REITs to Mortgage REITs in the USA, with further evidence finding that the two sectors are cointegrated.

In contrast to the literature that has examined the return behavior of REITs, few have examined volatility in the sector. Stevenson (2002) examined volatility spillovers within different REIT sectors and between REITs and the equity and fixed-income markets. The paper examined monthly data over the period 1975 to 2001. The study finds that volatility in Equity REITs has a significant influence on the other sub-sectors of the market and that a number of patterns emerge with regard to the influence of other asset classes. The primary results indicate that the REIT sector is generally influenced more strongly by volatility in small cap stocks and in firms classified as value stocks. These findings are not surprising given the average size of REITs and the fundamental nature of them. The S&P 500 has a mixed and inconsistent relationship with REITs, while there is no evidence of a positive relationship in volatility between the fixed income sector and Mortgage REITs. Devaney (2001) uses a GARCH-M model on monthly REIT data, primarily to examine the relationship between REIT volatility and interest rates. The paper finds significant influences on REIT returns from interest rate movements. The author does however find that in most cases the results for Equity REITs are not significant, with stronger findings reported for the Mortgage sector, as would to some extent be expected due to the nature of firms. Two recent working papers, Winniford (2003) and Najand & Lin (2004) both provide further evidence concerning the dynamics of daily volatility in the REIT sector. Najand & Lin (2004) utilize both a GARCH and GARCH-M model in their analysis of daily REIT volatility. The authors report evidence that would suggest that volatility shocks are persistent. Winniford (2003) concentrates on seasonaility in REIT volatility. The author finds strong evidence that volatility in Equity REITs does vary on a seasonal basis, with observed increased volatility in April, June, September, October and November.

The current paper extends previous studies in a number of ways. Firstly, in comparison to both Stevenson (2002) and Devaney (2001) it uses daily rather than monthly data. The use of daily data allows a deeper analysis of market based transmissions in volatility and also overcomes the problems inherent in using monthly REIT data due to the structural break in the early nineties. While a common criticism of using daily data is the noise contained within it, the use of higher frequency data allows an examination of whether previously reported results are stable over different data frequencies. In addition, the growth in the sector and especially the increased awareness from a broader class of investors is likely to result in an increase in daily volatility due to higher trading levels. These could result in changing dynamics in daily REIT volatility.

Secondly, in comparison to the previous studies of daily volatility the paper extends the analysis in a number of respects. While, for example, Najand & Lin (2004) do incorporate the general market into their model, they do not include other equity indices, and in particular, they do not examine the influence of value stocks. Likewise, Winniford (2003) does not include such equity sectors in his analysis. This is despite the strong empirical evidence in the literature concerning the linkages between REITs and the value sector. In addition the mainstream S&P 500 the current paper also analyses the interlinkages with the value and growth sectors as well as the NASDAQ which acts as a proxy for the technology sector. The paper also extends on both Winniford (2003) and Najand & Lin (2004) by not solely examining the Equity REIT sector, but also examines the volatility dynamics of the mortgage and hybrid sectors.

The paper is laid out as follows. The following section provides an initial description of the empirical approach used in the study. Section 3 describes the data used and provides preliminary statistics to assess the suitability of a GARCH approach in modeling daily REIT volatility. Section 4 contains the main empirical analysis, while the final section provides concluding comments.

**GARCH Models**

The main empirical analysis is undertaken in a GARCH (Generalized Autoregressive Conditional Heteroscedasticity) framework. GARCH models allow the simultaneous modeling of both the first and second moments of the return series' and provide a more efficient means of modeling time-series'. The use of ARCH based models allows us to examine the interlinkages between the different assets in terms of their second moment, effectively examining causal relationships in volatility. Conventional econometric time-series models assume that the variance of the error term is constant. This assumption of homoscedasticity is however often problematic in the analysis of financial time series', with the clustering of volatility being a prime example of a situation where this assumption may be violated.

The return generating process is modeled in a time-varying fashion:

$$R_t = bZ_t + \varepsilon_t \tag{1}$$

$$\varepsilon_t | I_{t-1} \sim N(0, H_t) \tag{2}$$

$$H_t = \alpha_0 + \sum \alpha_i \varepsilon_{t-i}^2 + \sum \beta_j H_{t-j}^2 \tag{3}$$

Where the mean is described by a first order VAR, and univariate volatility follows a GARCH (1, 1) process. The main advantage to the GARCH process proposed by Bollerslev (1986) is that it allows for lagged squared returns and volatility in the modeling process. A number of explanatory variables are also included in the mean and volatility specifications detailing the influences on the return generating process.

There are many possible extensions to this model that allow for some economic event affecting the volatility process. The most common being leverage effects modeled with an asymmetric term referred to as an (Exponential) EGARCH process. Black (1976) provided justification for examining whether the volatility measure is affected asymmetrically by the impact of positive and negative news, by documenting that volatility increased (decreased) as a result of periods of bad (good) news announcements. In essence, asset returns and future volatility are negatively related. A plethora of conditional stochastic volatility models are developed modeling this characteristic of which the Glosten et al (1993) model is chosen. This provides a more generalized version of the volatility process using the variance of returns that are constrained to having an asymmetric effect based on the general finding that negative news has a stronger effect on volatility than positive news. Here the additional variable, $S_i$, examines whether leverage is present in the volatility process.

$$H_t = \alpha_0 + \sum \alpha_i \varepsilon_{t-i}^2 + \sum \beta_j H_{t-j}^2 + \sum \gamma_i S_i \varepsilon_{t-i}^2 \qquad (4)$$

Where the asymmetric term, $S_i$, equals unity if $\varepsilon_{t-1}<0$ and zero otherwise[1].

**Data & Preliminary Analysis**

The data used in this paper consists of daily data for the period January 1999 to June 2003. The NAREIT daily indices are used throughout. Throughout the tests all three REIT sub-sectors are analyzed (Equity REITs, Mortgage REITs and Hybrid REITs). The All REIT Index was not examined in this study due both the differences inherent among the three sub-sectors and also due to the dominance of the Equity REIT sector. As of the end of 2003 Equity REITs accounted for 91% of the total market capitalization of the sector, with 144 out of a total of 171 listed REITs being classified as EREITs. Therefore, given this dominance the All REIT and Equity REIT indices provide largely identical results. Summary statistics for daily REIT returns are outlined in Exhibit 1. The characteristics are reasonably alike for the different REIT series with strongest variations being followed by Mortgage REITs. A similar positive daily average is recorded for each REIT series. However, both Mortgage and Hybrid REITs have daily standard deviation in excess of 1% with Mortgage REITs twice as risky as Equity REITs. Excess (positive) skewness and kurtosis is exhibited for each REIT series. In particular, the Mortgage REIT returns are extremely leptokurtotic. Normality is formally rejected for all series using the Kolmogorov-Smirnov test, with again, Mortgage REITs deviations being most pronounced[2].

{Insert Exhibits 1, 2 & 3}

Time series plots of daily returns are given in Exhibit 2. The pattern for Equity and Hybrid REITs follow a similar pattern, with returns constrained to a similar magnitude. All returns are time varying with volatility clusters. For instance the largest level of volatility occurred during end of 2002 for Equity REITs whereas Mortgage REITs incurred very large volatility during the middle of 1999 with returns in excess of 20% and – 10%. The other equity series follow a similar pattern to the Equity and Hybrid REIT series' and are displayed in Exhibit 3. The range of returns is reasonably similar between the REIT and S&P series' although the small capitalization companies captured by the S&P Value index are capable of larger absolute returns. The NASDAQ is also more volatile than the REIT series.

{Insert Exhibits 4 & 5}

An initial examination of the dynamics of REIT returns and volatility can be undertaken by assessing the dependencies present in the return and volatility series through the autocorrelation function (ACF). Exhibits 4 and 5 display the ACF over 30 lags, with squared returns used to detail characteristics of the volatility series. Similar findings in line with financial time series in general are identified with low persistence in returns being contrasted by relatively strong persistence in the volatility series. This finding is associated with the volatility clustering property of financial time series. The strong serial correlation of volatility indicates the existence of ARCH effects and validities the application of GARCH related processes. Notwithstanding the significant dependency of returns in the first lag due to non-synchronous trading, there is a general lack of significant autocorrelation in all returns series. The high first order autocorrelation reported is to some extent expected due to the small average relative size of REITs and the average level of daily trading in the sector. The persistence of this series' volatility follows a similar pattern to financial time series in general with strong significant dependence decreasing slowly and remaining over the first ten lags and thereafter being insignificant. Persistence in volatility is weakest for Mortgage REITs although it does record the largest single autocorrelation estimate.

{Insert Exhibit 6}

Summary statistics for daily REIT volatility are outlined in Exhibit 6. Daily conditional standard deviations are analyzed from fitting the GARCH (1, 1) model[3]. The characteristics of the returns series are followed by the volatility series with similar properties being presented in general. Differences occur between returns and volatility in terms of magnitude of values, with for example, the non-normality of the volatility series being much stronger than the returns series. Again, stronger deviations across the series are exhibited by Mortgage REITs. For instance the volatility of volatility (standard deviation) measure is approximately 0.2 for all series with the exception of Mortgage REITs with a figure of almost 1% daily. Also the excess skewness and kurtosis is strongest for this series vis-à-vis the other REITs. These estimates dwarf the values for the returns series generating highly significant Kolmogorov-Smirov values. Finally the volatility of Mortgage REITS exhibits the most extreme values compared to the other series, with relatively large 3$^{rd}$ quartile and maximum values.

{Insert Exhibits 7 & 8}

Daily-standardized returns and conditional volatility (% Standard deviation) are plotted for all series in Exhibits 7 and 8. The estimates are again from the GARCH model. The objective of the standardized returns, returns scaled by conditional volatility, is to indicate the ability of the GARCH model to capture the time-varying dynamics of the respective returns series. The plots suggest that the GARCH model is quite effective, for example, dampening the very extreme values of the Mortgage REITs returns. As with the return series the conditional volatility estimates for the Equity and Hybrid sectors are similar, with the highest reported figures of volatility being observed in 2002. In contrast daily volatility for Mortgage REITs showed a very different pattern with excessive volatility occurring during 1999. The REITs, with the exception of the Mortgage series, followed a reasonably similar pattern to the S&P equity series, although the latter set of series incurred higher levels of volatility at their extremes. The plots suggest that investors in the REIT markets are influenced by activity in the general equity market. The pattern for new technology equities given by the NASDAQ series suggests a slightly different pattern with largest volatility during 2000 during the final year of the technology boom.

**Empirical Analysis of REIT Volatility**

Exhibits 9 and 10 report the results from the fitting of the GARCH (1,1) and EGARCH (1,1) models respectively. The estimations for each series are for the entire sample period and are made using maximum likelihood methods of the conditioning variables. In general the findings are in line with expectations for the GARCH parameters with significant parameters. The set of explanatory variables included for analysis for each REIT index are the other REIT series returns, a range of other equity sectors and a number of additional variables. The influence of the equity markets is proxied by the inclusion of a number of alternative equity market indices. The S&P500 Composite is used as a proxy of the overall market. New economy firms are proxied by the NASDAQ Composite, while the S&P mid-cap value and growth indices are also incorporated into the analysis. The rationale behind the inclusion of value indices in particular concerns the characteristics of REITs. Most REITs are mid and small cap stocks and due to the nature of them generally have more in common with value firms, with relatively high asset value to market value, than growth stocks. This has been noted in a number of studies, including Chiang & Lee (2002), who found using Style Analysis that EREITs can be classified as a combination of value stocks and t-bills. In addition, the results of Stevenson (2002) in his analysis of volatility spillovers using monthly REIT data would support this view, finding that value stocks were more significant in terms of volatility transference than the large cap S&P500 and NASDAQ or growth indices. The other variables included in the models are the one-month US Treasury Bill, and the US CPI. Seasonal effects are accounted for by including a Monday Variable, while dummy variables for the market crashes of April 2000 and September 2001 are also included. The inclusion of these dummy variables allows for the possibility that some form of structural break occurred around these two events for concerning the interaction between REITs and other asset classes. The performance of the S&P500 series is a strong explanatory variable for the Equity REIT return generating processes. In contrast, the impact of the respective REIT returns series on each other is negligible with the exception of the Hybrid REIT series, which is influenced by Equity REIT returns. Whilst the US Treasury Bill, measuring short-term money rates, is an important driving force for Equity REITs, the other macroeconomic variable, the CPI, has negligible influence for the REIT series. All variables included have no explanatory power for the Mortgage REITs returns.

{Insert Exhibits 9 & 10}

Turning to the volatility model, the GARCH parameters are significant and in line with previous daily studies, with the strong influence of past squared returns and past volatility recorded. The asymmetric leverage term in the EGARCH specification has mixed results, with it not being influential for either the equity or hybrid sectors. In contrast however, strong negative leverage effects are recorded for the

Mortgage REITs. For all the series, the sum of the volatility parameters implies stationarity due to their summing too less than unity. Generally Equity REIT volatility influences the other REIT markets but not vice-versa. Furthermore, volatility in all the REIT series are affected differently by activity in the other equity markets such as the S&P Value affecting the Mortgage sector but not the series from fitting the GARCH (1,1) model. In this GARCH model money market activity through the US Treasury Bill is influential on Equity REITs whereas this does not occur for the EGARCH specification. Returns on the NASDAQ are the only important variable influencing the volatility of Hybrid REITs. The results are notably not as strong as those reported in the monthly analysis of Stevenson (2002) in terms of the influence of value and small/mid cap stocks. In only one case, that of Mortgage REITs with the GARCH specification, are significant spillover effects reported for the S&P Value Index. While this finding appears initially to be intuitive it does require explanation. Due to the differing characteristics of the Equity and Mortgage sectors the intuitive fundamental linkages between value stocks and Mortgage REITs are not as strong as with the equity sector. It is perhaps that the results reported here are due to the exact time period covered. Following the collapse of the technology boom in 2000 and the resulting poor performance in the growth sector, portfolio managers diverted funds into alternative sectors, including value stocks, but also the fixed income market. It is therefore perhaps due to this increase in the flow of funds into the fixed income and money markets which in turn would influence Mortgage REITs. In contrast to these findings, the large cap indices, and especially the S&P 500 provide more consistent influence on REIT volatility. Likewise, the interlinkages between the three REIT sub-markers are not as strong as reported by Stevenson (2002). This would indicate that the use of daily data provides a contrasting picture compared to the analysis of monthly data. Monthly data would appear to allow more time for the more substantial and intuitive relationships to come to the fore. However, on a daily basis it would appear that these relationships are less clear and that the markets tend to follow broad sentiment trends with less fundamental relationships at play. Therefore, the inter-relationships between REIT sectors and between them and related sectors, such as value stocks, is to some extent masked at the higher frequency daily data, with the general market being more influential.

A spectrum of diagnostic tests from fitting the GARCH model is displayed in Exhibit 11. The Ljung-Box test examines the residuals and squared residuals and finds negligible serial dependence. The Engle (1982) LM test for ARCH of order $p$ tests the null of zero slopes in the regression:

$$y_{i,t}^2 = \phi_0 + \sum_{m=1}^{p} \phi_m y_{i,t-m}^2 + u_{i,t} \qquad (5)$$

The test is performed as $T \cdot R^2$ where T represents the sample size. Similarly, in all cases the null of ARCH effects remaining after fitting the GARCH specification was rejected at all usual levels of confidence. A common finding in the volatility modeling literature, known as the leverage effect or asymmetric volatility, is that negative shocks cause more volatility than positive shocks of equal magnitude. The asymmetry is first examined by reporting the Engle and Ng (1993) tests for size and sign bias applied to the series returns. Define $N_{i,t}$ as in indicator dummy that takes the value 1 if $y_{i,t} < 0$ and zero otherwise. The test for sign bias is based on the significance of $\phi_1$ in:

$$y_{i,t}^2 = \phi_0 + \phi_1 N_{i,t-1} + u_{i,t} \tag{6}$$

If positive and negative innovations have differing impacts on the conditional variance of $y_{i,t}$, then $\phi_1$ will be statistically significant, which is not supported for the residual series. It may also be the case that the source of the bias is caused not only by the sign, but also the magnitude of the shock. The negative size bias test is based on the significance of the slope coefficient $\phi_2$ in:

$$y_{i,t}^2 = \phi_0 + \phi_2 N_{i,t-1} y_{i,t-1} + u_{i,t} \tag{7}$$

Likewise, defining $P_{i,t} = 1 - N_{i,t}$, a similar test may be performed for positive size bias with no sign bias indicated. Finally, the Engle and Ng (1993) joint test for asymmetry in variance is based on the regression:

$$y_{i,t}^2 = \phi_0 + \phi_1 N_{i,t-1} + \phi_2 N_{i,t-1} y_{i,t-1} + \phi_3 P_{i,t-1} y_{i,t-1} + u_{i,t} \tag{8}$$

Significance of the parameter $\phi_1$ indicates the presence of *sign bias*. That is, positive and negative realizations of shocks to the process affect future volatility differently to the prediction of the model. Similarly significance of $\phi_2$ or $\phi_3$ would suggest *size bias*, where not only the sign, but also the magnitude of innovation in growth is important. A joint test for sign and size bias, based upon the Lagrange Multiplier Principle is performed. The results indicate that the residuals series exhibit no leverage effects.

{Insert Exhibit 11}

**Conclusion**

This paper has examined the relationships between different REIT sectors in the context of univariate GARCH and EGARCH models. The paper has attempted to examine both the causes and properties of volatility in REIT returns. The results highlight the linkages between the different sub-markets and also the S&P 500 and other mainstream stock indices. The return and volatility findings are in line with previous studies to have examined longer horizons and also provide evidence that GARCH based models are suitable in the analysis of REIT volatility. The use of daily data does provide a number of contrasting volatility spillover findings to those reported by Stevenson (2002) who analyzed the monthly NAREIT indices. In particular it would appear that the more fundamentally based and intuitive results reported in that study are harder to capture when the higher frequency daily data is used. The relationship with value stocks is weakened considerably, while the influence of the large cap sector is enhanced. The exact causes of these diverging results is not however clear. One alternative cause is that the additional noise contained in daily data results in broad market sentiment playing a more significant role, with the more intuitive and perhaps fundamental relationships being masked at this higher frequency. The second alternative is that we are actually observing a structural change in the relationship between REITs and other sectors due to increased investor awareness and investment in the sector. Given that most previous studies have largely relied on long-term monthly data, the use of this short-term higher frequency data do open up the possibility that such a break has occurred, in many respects perhaps similar to that observed in the early nineties.

**Exhibits**

## Exhibit 1: Descriptive Statistics of Daily Returns Series

|  | Equity REITs | Mortgage REITs | Hybrid REITs |
|---|---|---|---|
| **Min** | -3.681 | -17.974 | -4.258 |
| **1st quartile** | -0.408 | -0.532 | -0.565 |
| **Median** | -0.009 | 0.085 | 0.000 |
| **3rd quartile** | 0.405 | 0.692 | 0.611 |
| **Max** | 4.610 | 21.639 | 4.969 |
| **Mean** | 0.014 | 0.021 | 0.011 |
| **Std Deviation** | 0.756 | 1.562 | 1.031 |
| **Skewness** | 0.135 | 0.473 | 0.017 |
| **Kurtosis** | 4.129 | 51.642 | 2.238 |
| **Normality** | 0.057 | 0.122 | 0.049 |

Note: This table gives the summary statistics for daily REIT returns. Excess skewness and kurtosis are detailed with critical values of 0. It also presents the results of the Kolmogorov-Smirnov test for normality. All skewness, kurtosis and normality test coefficients are significant.

**Exhibit 2: Time Series Plots for Daily REIT Returns**

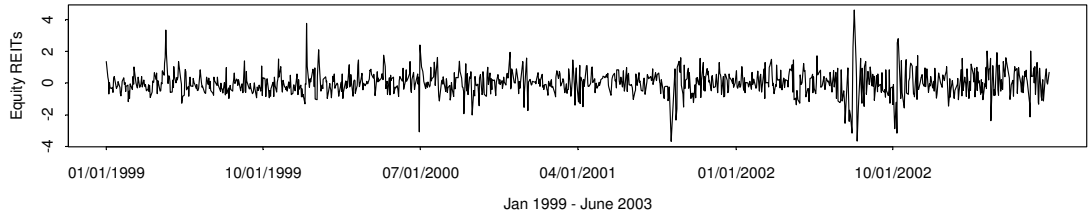

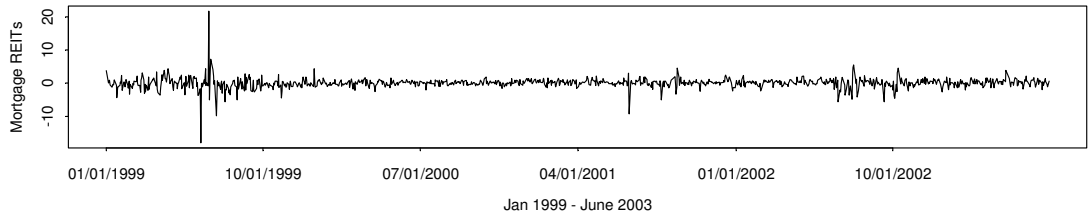

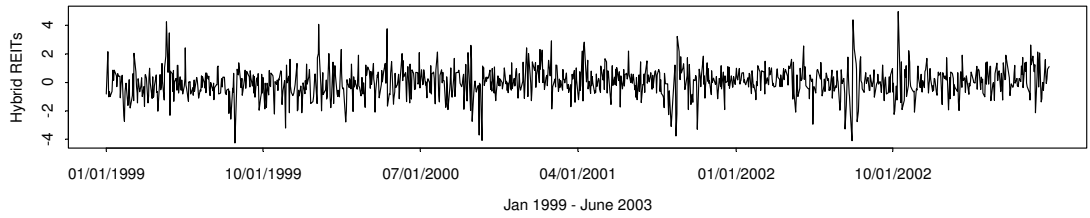

**Exhibit 3: Time Series Plots for Daily Equity Returns**

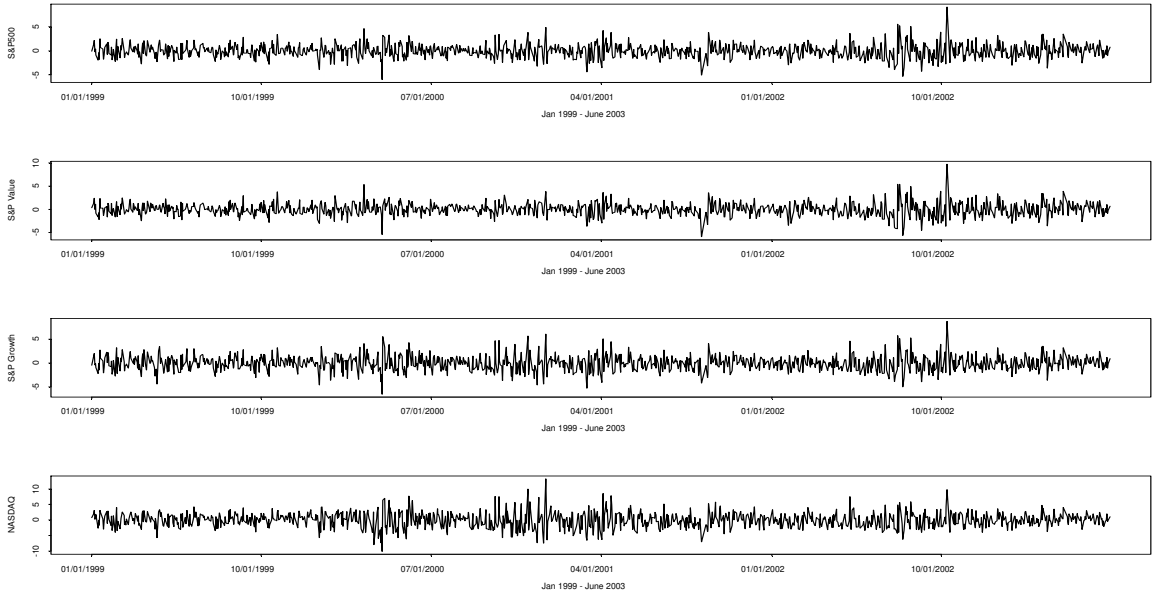

**Exhibit 4: Autocorrelation Plots for Daily REIT Returns**

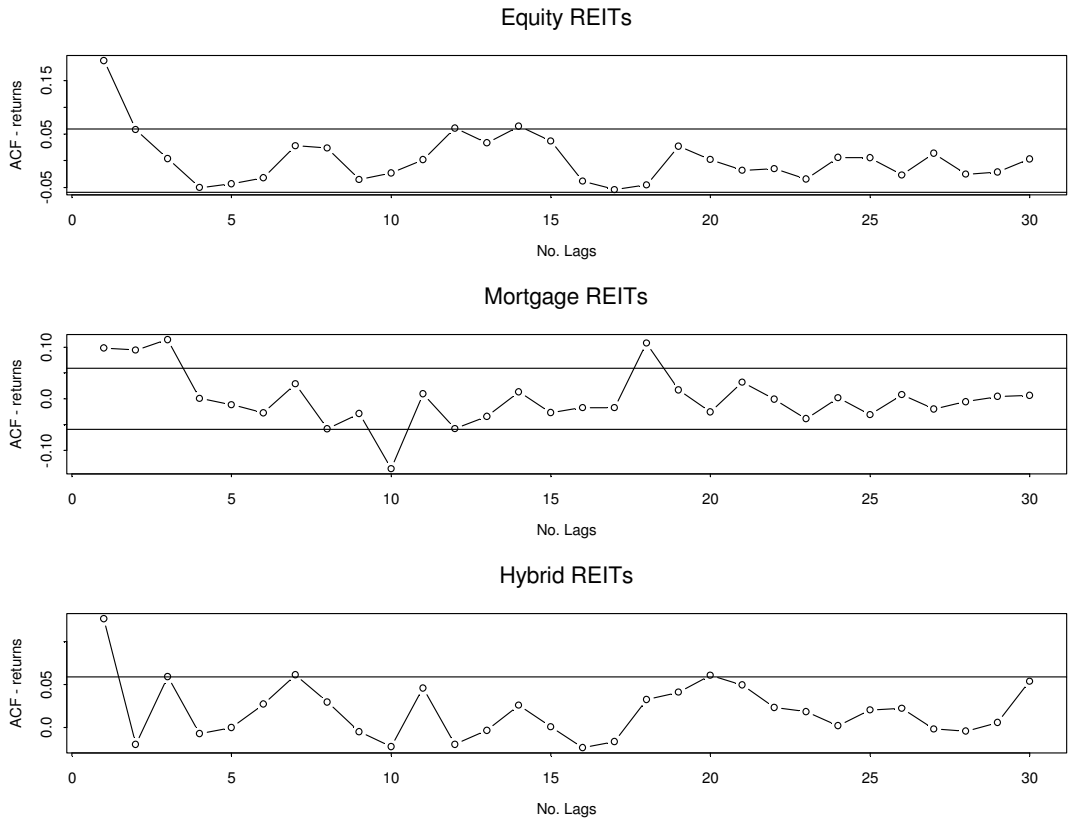

Notes: All plots include confidence bands measured by ± 1.96/√T so significance occurs at ± 0.059 and these are imposed where appropriate.

**Exhibit 5: Autocorrelation Plots for Daily REIT Volatility**

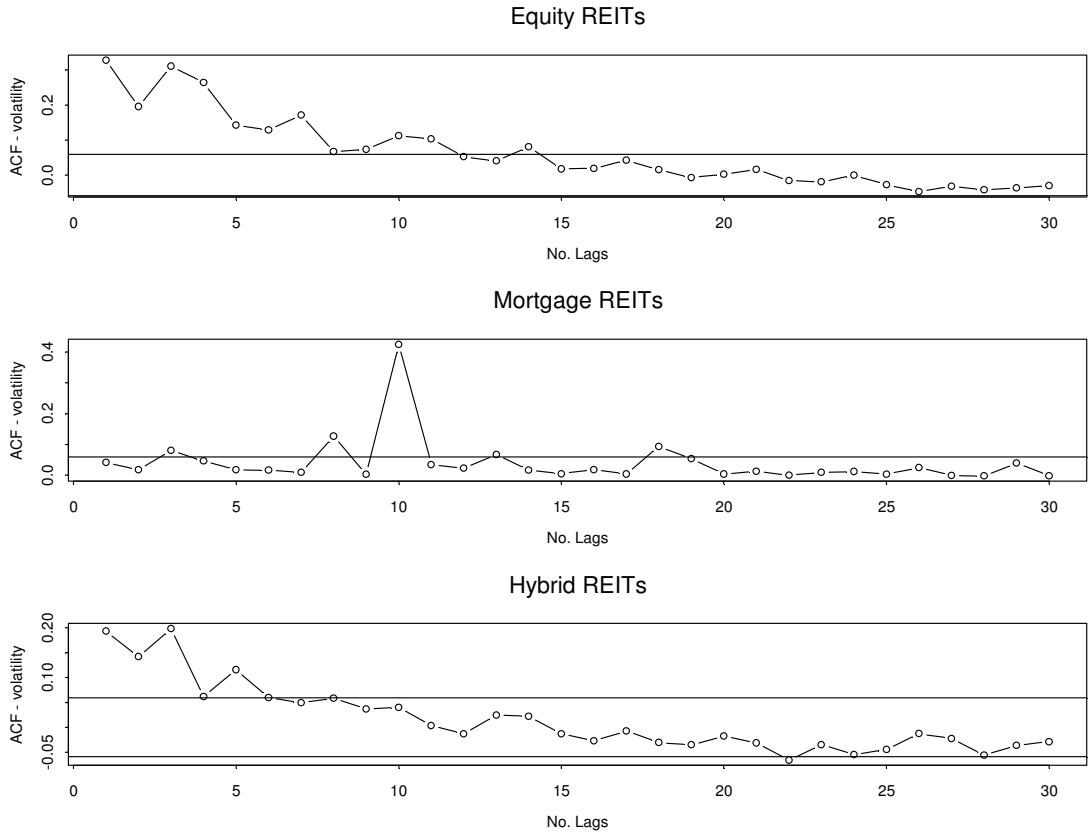

Notes: All plots include confidence bands measured by ± 1.96/√T so significance occurs at ± 0.059 and these are imposed where appropriate.

**Exhibit 6: Summary statistics of daily REIT volatility**

|  | Equity REITs | Mortgage REITs | Hybrid REITs |
|---|---|---|---|
| Min | 0.519 | 0.646 | 0.722 |
| 1$^{st}$ quartile | 0.582 | 0.879 | 0.856 |
| Median | 0.644 | 1.048 | 0.940 |
| 3$^{rd}$ quartile | 0.759 | 1.460 | 1.058 |
| Max | 2.268 | 8.689 | 2.404 |
| Mean | 0.710 | 1.338 | 0.990 |
| Std Deviation | 0.214 | 0.902 | 0.211 |
| Skewness | 2.955 | 4.076 | 2.489 |
| Kurtosis | 11.603 | 21.793 | 9.087 |
| Normality | 0.188 | 0.227 | 0.141 |

Note: Exhibit 6 gives the summary statistics for daily REIT volatility. Volatility is obtained from fitting the GARCH (1, 1) specification. Excess skewness and kurtosis are detailed with critical values of 0. It also presents the results of the Kolmogorov-Smirnov test for normality. All skewness, kurtosis and normality test coefficients are significant.

**Exhibit 7: Time Series Plots for Daily Standardized REIT Returns**

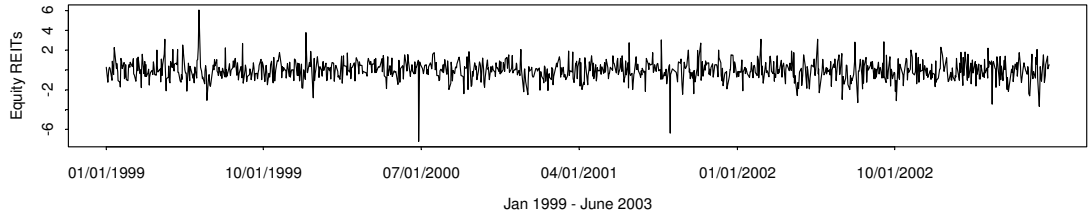

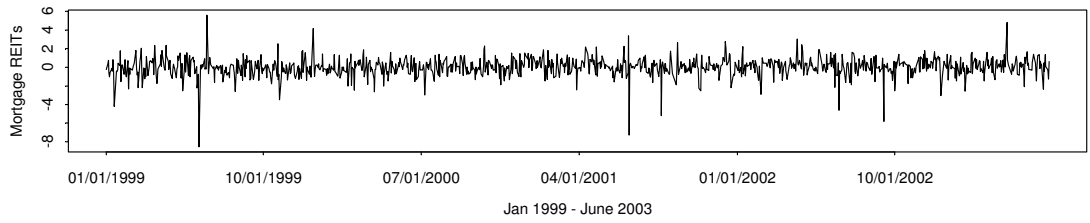

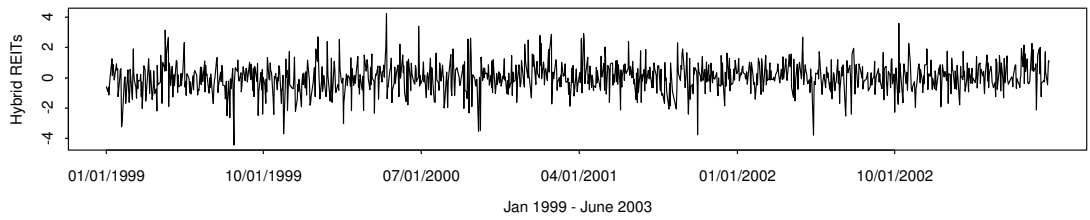

**Exhibit 8: Time Series Plots for Daily Conditional Volatility for Return Series**

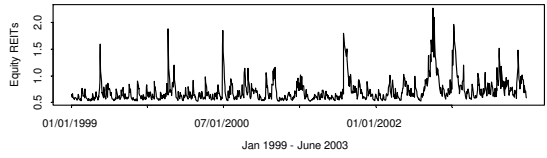
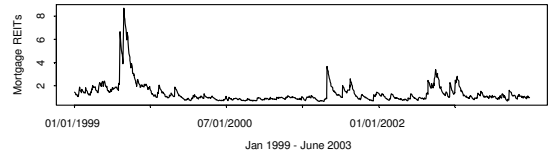
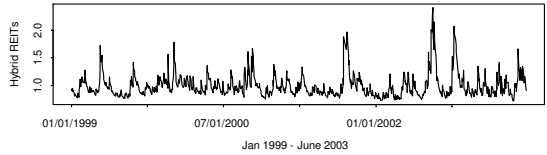
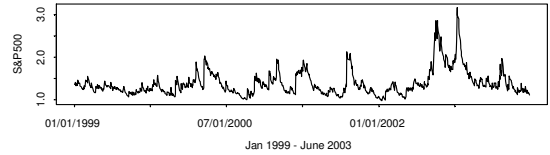
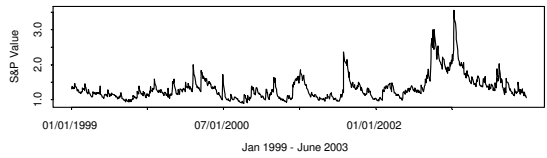
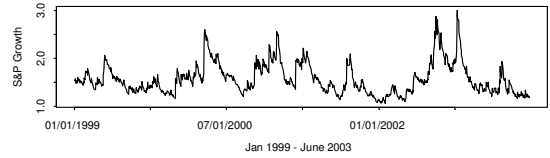
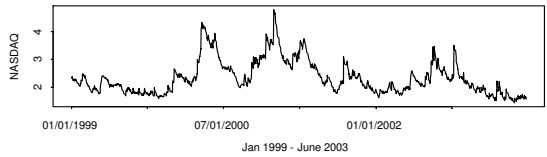

**Exhibit 9: Conditional Modeling of Daily Returns Series, GARCH (1,1)**

|  | Equity REITs | | Mortgage REITs | | Hybrid REITs | |
|---|---|---|---|---|---|---|
|  | Coefficient | p-value | Coefficient | p-value | Coefficient | p-value |
| **Panel A: Conditional Returns** | | | | | | |
| Constant | 0.000 | 0.133 | 0.001 | 0.000 | 0.000 | 0.482 |
| Equity REITs | 0.109 | 0.001 | 0.041 | 0.324 | 0.100 | 0.019 |
| Mortgage REITs | 0.007 | 0.601 | 0.135 | 0.000 | 0.004 | 0.780 |
| Hybrid REITs | 0.034 | 0.088 | 0.033 | 0.265 | 0.061 | 0.075 |
| Monday | -0.001 | 0.134 | -0.001 | 0.396 | 0.000 | 0.752 |
| April | 0.004 | 0.000 | -0.002 | 0.381 | 0.001 | 0.647 |
| September | -0.004 | 0.501 | 0.003 | 0.692 | -0.006 | 0.404 |
| S&P500 | -0.672 | 0.027 | 0.164 | 0.753 | -0.810 | 0.265 |
| Treasury Bill | 0.011 | 0.001 | 0.008 | 0.459 | 0.008 | 0.510 |
| CPI | 0.013 | 0.677 | 0.003 | 0.935 | 0.024 | 0.607 |
| NASDAQ | 0.053 | 0.000 | -0.047 | 0.025 | -0.010 | 0.685 |
| S&P Value | 0.591 | 0.000 | 0.188 | 0.460 | 0.630 | 0.070 |
| S&P Growth | 0.255 | 0.109 | 0.008 | 0.977 | 0.462 | 0.229 |
| **Panel B: Conditional Volatility** | | | | | | |
| $a_1$ | 0.000 | 0.000 | 0.000 | 0.321 | 0.000 | 0.029 |
| $a_2$ | 0.226 | 0.000 | 0.288 | 0.000 | 0.079 | 0.000 |
| $b_1$ | 0.359 | 0.000 | 0.752 | 0.000 | 0.841 | 0.000 |
| Equity REITs |  |  | 0.001 | 0.097 | -0.001 | 0.004 |
| Mortgage REITs | 0.000 | 0.000 |  |  | 0.000 | 0.343 |
| Hybrid REITs | 0.000 | 0.219 | -0.001 | 0.008 |  |  |
| Monday | 0.000 | 0.263 | 0.000 | 0.086 | 0.000 | 0.232 |
| April | 0.000 | 0.331 | 0.000 | 0.512 | 0.000 | 0.248 |
| September | 0.000 | 0.471 | 0.000 | 0.761 | 0.000 | 0.338 |
| S&P500 | 0.001 | 0.646 | -0.013 | 0.017 | 0.012 | 0.070 |
| Treasury Bill | 0.000 | 0.048 | 0.000 | 0.316 | 0.000 | 0.207 |
| CPI | 0.000 | 0.525 | 0.000 | 0.203 | 0.000 | 0.438 |
| NASDAQ | 0.000 | 0.560 | 0.000 | 0.076 | 0.000 | 0.002 |
| S&P Value | -0.001 | 0.571 | 0.007 | 0.015 | -0.005 | 0.088 |
| S&P Growth | 0.000 | 0.761 | 0.006 | 0.030 | -0.007 | 0.047 |

Notes: Results for conditional mean and volatility with explanatory variables are reported as described in text. Marginal significance levels using Bollerslev-Wooldridge standard errors are displayed by parentheses. * denotes significance at the 5% level.

**Exhibit 10: Conditional Modeling of Daily Returns Series, EGARCH (1,1)**

|  | Equity REITs | | Mortgage REITs | | Hybrid REITs | |
|---|---|---|---|---|---|---|
|  | Coefficient | p-value | Coefficient | p-value | Coefficient | p-value |
| **Panel A: Conditional Returns** | | | | | | |
| Constant | 0.000 | 0.380 | 0.001 | 0.001 | 0.000 | 0.344 |
| Equity REITs | 0.110 | 0.000 | 0.037 | 0.337 | 0.107 | 0.012 |
| Mortgage REITs | -0.007 | 0.536 | 0.152 | 0.000 | 0.009 | 0.533 |
| Hybrid REITs | 0.020 | 0.317 | 0.032 | 0.303 | 0.041 | 0.221 |
| Monday | -0.001 | 0.151 | -0.001 | 0.360 | 0.000 | 0.814 |
| April | 0.003 | 0.003 | -0.001 | 0.434 | 0.001 | 0.605 |
| September | -0.005 | 0.131 | 0.002 | 0.692 | -0.006 | 0.297 |
| S&P500 | -0.917 | 0.001 | -0.306 | 0.505 | -0.620 | 0.378 |
| Treasury Bill | 0.010 | 0.194 | 0.005 | 0.613 | 0.007 | 0.551 |
| CPI | 0.028 | 0.315 | 0.009 | 0.833 | 0.018 | 0.688 |
| NASDAQ | 0.046 | 0.001 | -0.046 | 0.044 | -0.014 | 0.580 |
| S&P Value | 0.683 | 0.000 | 0.424 | 0.062 | 0.532 | 0.115 |
| S&P Growth | 0.413 | 0.003 | 0.241 | 0.307 | 0.376 | 0.310 |
| **Panel B: Conditional Volatility** | | | | | | |
| $a_1$ | -1.085 | 0.000 | -0.181 | 0.001 | -0.547 | 0.000 |
| $a_2$ | 0.312 | 0.000 | 0.384 | 0.000 | 0.138 | 0.000 |
| $b_1$ | 0.896 | 0.000 | 0.984 | 0.000 | 0.945 | 0.000 |
| Leverage | 0.037 | 0.649 | -0.105 | 0.037 | -0.144 | 0.322 |
| Equity REITs |  |  | 3.649 | 0.295 | -9.476 | 0.000 |
| Mortgage REITs | 0.568 | 0.754 |  |  | -1.604 | 0.193 |
| Hybrid REITs | -0.904 | 0.724 | -5.478 | 0.002 |  |  |
| Monday | 0.141 | 0.235 | 0.371 | 0.000 | 0.145 | 0.188 |
| April | 0.033 | 0.697 | -0.071 | 0.469 | 0.097 | 0.051 |
| September | 0.629 | 0.054 | 0.136 | 0.587 | 0.303 | 0.220 |
| S&P500 | 278.441 | 0.001 | -140.372 | 0.199 | 39.490 | 0.629 |
| Treasury Bill | 1.752 | 0.321 | 1.524 | 0.189 | 1.064 | 0.421 |
| CPI | -1.695 | 0.697 | 5.758 | 0.095 | -3.949 | 0.253 |
| NASDAQ | 3.246 | 0.128 | 2.118 | 0.273 | 4.982 | 0.003 |
| S&P Value | -133.963 | 0.001 | 65.491 | 0.224 | -13.749 | 0.723 |
| S&P Growth | -148.970 | 0.001 | 73.951 | 0.182 | -30.925 | 0.469 |

Notes: Results for conditional mean and volatility with explanatory variables are reported as described in text. Marginal significance levels using Bollerslev-Wooldridge standard errors are displayed by parentheses. * denotes significance at the 5% level.

**Exhibit 11: Diagnostics Tests of Conditional Modeling**

| | All REITs | | Equity REITs | | Mortgage REITs | | Hybrid REITs | |
|---|---|---|---|---|---|---|---|---|
| | Coefficient | p-value | Coefficient | p-value | Coefficient | p-value | Coefficient | p-value |
| Q(24) | 30.5468 | 0.16728 | 30.6984 | 0.16267 | 17.3952 | 0.83132 | 22.5494 | 0.54651 |
| $Q^2$(24) | 13.2403 | 0.96199 | 13.139 | 0.96377 | 68.2574 | 0 | 22.4561 | 0.55207 |
| ARCH(24) | 0.5383 | 0.96647 | 0.5279 | 0.97038 | 3.235 | 0 | 0.9334 | 0.55554 |
| N-SIGN | 0.0925 | 0.92632 | 0.2527 | 0.80059 | 1.266 | 0.20581 | -0.5322 | 0.59467 |
| N-SIZE | 0.2106 | 0.83327 | 0.1419 | 0.88721 | 0.2029 | 0.83924 | -0.4287 | 0.66822 |
| P-SIZE | 0.3601 | 0.71882 | 0.4844 | 0.6282 | 0.3309 | 0.74077 | 0.7003 | 0.48393 |
| JOINT | 0.0844 | 0.96855 | 0.0877 | 0.96682 | 0.5986 | 0.61598 | 0.5957 | 0.61788 |

Notes: Q(24) is a Ljung-Box test on the residual series whereas $Q^2$(24) is the Ljung-Box test on the squared residuals. ARCH(24) is the Engle (1981) LM test for up to twenty fourth order ARCH. N-Sign, N-Size, P-Size and Joint are Engle-Ng (1993) LM tests for asymmetry in variance. Marginal significance levels displayed in parenthesis.

**Endnotes:**

[1] An alternative to the univariate approach used in this study is to use a multivariate GARCH framework. This also allows examination of the time-varying correlations between the respective assets. A companion paper to this by the authors extends the analysis contained in this study to examine these issues.

[2] The stationarity of the three indices was tested using the Augmented Dickey Fuller (ADF) unit root test. The findings, which are available from the authors on request, are consistent with past studies modeling equity and REIT series', namely that the price data does not accept the hypothesis of stationarity. Following convention, the price series' are first differenced resulting in stationary series and avoiding avoid spurious conclusions. Further analysis will concentrate on these returns series.

[3] The inferences from fitting the EGARCH model are similar and are available on request.